# Radiative cooling capacity on Earth

Cunhai Wang,[1,#,*] Hao Chen,[1,#] Yanyan Feng,[2,#] Ziming Cheng,[3] Jingchong Liu,[4,*] Fuqiang Wang[3,*]

[1]School of Energy and Environmental Engineering, University of Science and Technology Beijing, Beijing 100083, China

[2]College of Mechanical and Transportation Engineering, China University of Petroleum Beijing, Beijing 102249, China

[3] School of Energy Science and Engineering, Harbin Institute of Technology, Harbin, 150001, China

[4] School of Chemistry and Biological Engineering, University of Science and Technology Beijing; Beijing, 100083, China

[#]Contributed equally.

*Correspondence: wangcunhai@ustb.edu.cn; jchliu@ustb.edu.cn; wangfuqiang@hitwh.edu.cn

**ABSTRACT:** By passively dissipating thermal emission into the ultracold deep space, radiative cooling (RC) is an environment-friendly means for gaining cooling capacity, paving a bright future for global energy saving and carbon dioxide reduction. However, assessing the global RC capacity at the day-to-annual scale remains challenging as the RC capacity significantly depends on geographic and environmental conditions. To our knowledge, no analysis of global RC capacity has been reported. Herein, we show the distribution of RC capacity on Earth by establishing a precise assessment model for evaluating the performance of a radiative cooler. Our assessment is comprehensively validated against experimental data and extended to elucidate the capacity of representative broadband and selective cooler. We also categorize the global RC capacity into five representative regions based on the year-round cooling power. Our assessment can inform trade-offs between design and practical application for the RC systems, alongside promoting RC-based technologies to tackle worldwide energy and environment challenges.

## 1. Introduction

Radiating thermal energy in electromagnetic waves, referred to as thermal radiation, is spontaneous and ubiquitous for terrestrial objects. By dissipating thermal radiation with a wavelength between 8 – 13 μm into the ultracold deep space (~ 3 K) through the high transparent atmosphere window, a sky-facing surface on Earth can be self-cooled, even to a sub-ambient temperature. This phenomenon is widely known as passive RC[1–6]. The technology of RC is free of nonrenewable fuels, all-day continuous, and eco-friendly. It has recently sparked increasing research interests[7–12] and paves a promising way to overcome worldwide issues such as global warming[13] and water scarcity[14].

The capacity of RC is achieved from the fact that a majority part of the thermal radiation emitted from a terrestrial surface can escape from the atmosphere window within the 8 – 13 μm wavelength range. Therefore, the RC performance is determined by the spectral characteristics of the inclusion atmosphere between the terrestrial radiating surface and the space cold sink[15]. In current RC applications, the model of constant-fixed effective atmospheric transmissivity (EAT)[16–18] is usually adopted to evaluate the cooling performance of a RC cooler[19-21]. However, using the EAT may lead to strong deviations in the RC performance[22], as the atmosphere spectrum differs in different locations, seasons, and even at different day periods[23–25].

Previous reports[26–28] indicate that atmospheric transmittance is closely related to the moisture in the atmosphere and can be approximately obtained through an environmental parameter of local real-time precipitable water value (PWV). Zhao et al.[29] studied the PWV effects on a high-performance RC module located in Boulder, Colorado. Based on the hourly meteorological data from China's observation stations, Zhu et al.[30] calculated the atmospheric spectral emissivity under different PWV values and then predicted the RC potential in China. In practice, however, with the local climate changing with locations and seasons, the PWV is neither accessible from existing databases nor readily measured[31]. Thus, the cooling power prediction model developed for one scenario cannot be applied to others. Limited studies have examined the RC capacity at the city or seasonal scales[32]. Assessing the global RC capacity at the seasonal and annual scales is essential to evaluate the performance of radiative coolers, improve their applicability and reliability, and promote RC technologies in real contexts. But to date, such capacity assessment is still lacking.

In this work, we evaluate and categorize the global RC power. We first calculated the worldwide PWV based on realistic meteorological data. The PWV data were then applied to obtain the spectral distribution of the atmospheric transmissivity. Afterward, we conducted long-time experimental tests and fully verified the accuracy and utility of our model for assessing RC performance. Then, we calculated the global distribution of RC power. The cooling capacities obtained by using broadband and spectral selective coolers were compared. Finally, based on the cooling power around the world, we depicted the global RC map and classified it into five typical regions, paving practical guidance for assembling engineering applications driven by passive RC.

## 2. Establishing the assessing model

The workflow of the assessing model is illustrated in **Fig. 1A**. We first grabbed the hourly meteorological data of 1,986 observation stations (**Fig. 1B**) around the world.[33] By using the hourly meteorological data as inputs, we calculated the PWV via the basic theory of global positioning system-based PWV estimation [34,35] (see PWV model in Methods). Then, we generated the spectral transmissivity (**Fig. 1C**) of the atmosphere at different PWV.[36] The varying atmospheric transparency with PWV indicates that the cooling performance of a radiative cooler will be closely related to its location and emission characteristics. Here, two kinds of coolers (with spectral emissivity distributions presented in **Fig. 1C**) are considered for studying RC capacity. One is a broadband

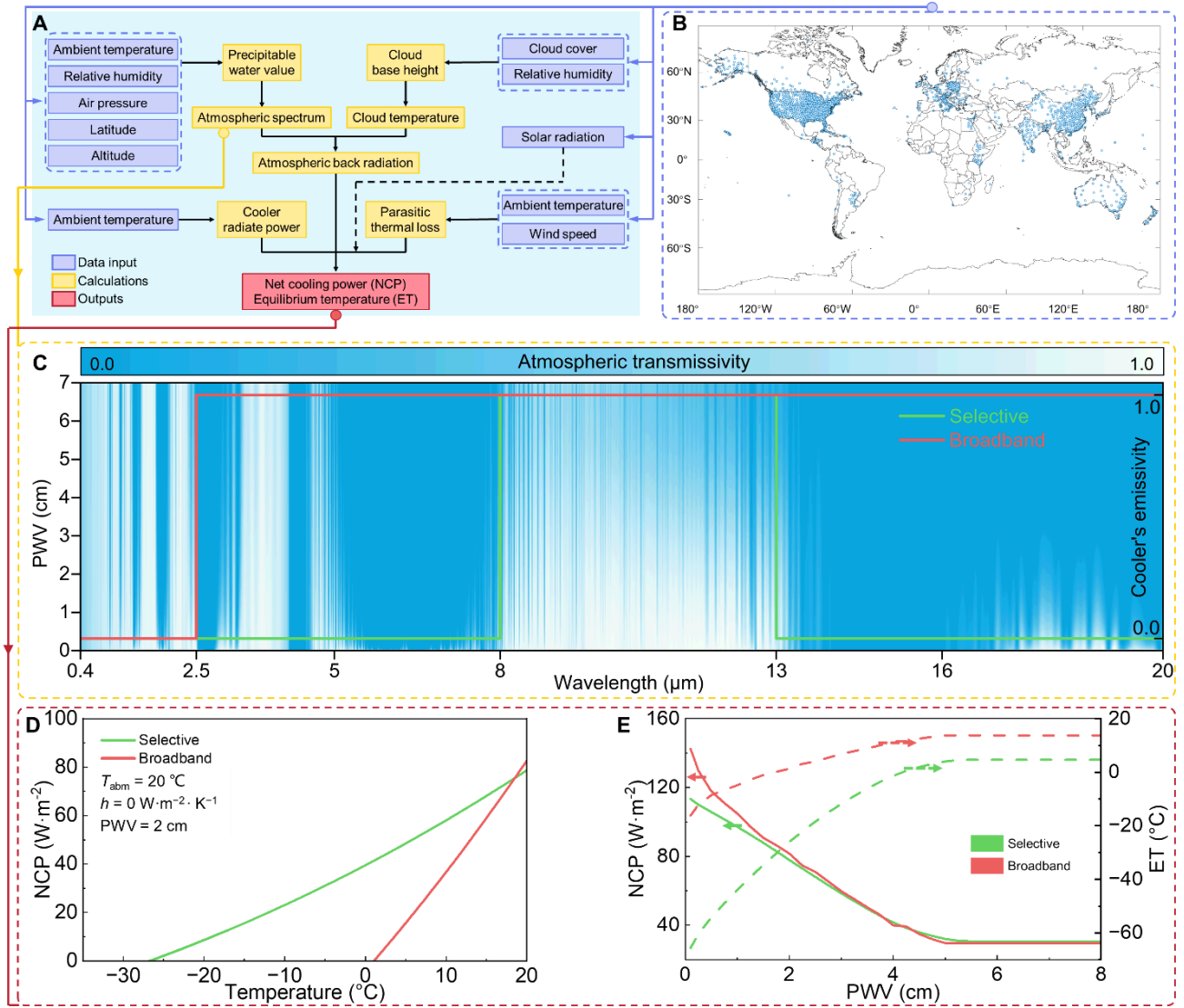

**Figure 1. Illustration of the assessing model.**

(**A**) Workflow of the assessing model for evaluating global radiative cooling (RC) capacity.

(**B**) Geographic distribution of the meteorological stations.[33]

(**C**) Transmissivity of the atmospheric at different PWVs and the emissivity of ideal selective or broadband radiative coolers.

(**D**) Net cooling power (NCP) of the two kinds of radiative coolers at different temperatures.

(**E**) NCP and equilibrium temperature (ET) of the two kinds of radiative coolers at different PWV.

radiative cooler with a unit emissivity of $\varepsilon_{e,\,2.5-20} = 1.0$ within the wave range of 2.5 – 20 μm. The other one is a spectral selective radiative cooler with a unit emissivity of $\varepsilon_{e,\,8-13} = 1.0$ within the 8 – 13 μm atmosphere window spectrum. Both coolers are non-emitting/absorbing within other spectrums.

We then calculated the atmospheric back radiation by using the atmospheric spectrum and cloud temperature. Assuming an optimal condition without any thermal loss via parasitic conduction and convection, we present in **Fig. 1D** the net cooling power (NCP) versus the cooler's temperature at a PWV of 2.45 cm (to be consistent with referenced data) and 20 °C ambient temperature ($T_{amb}$), which show excellent consistency with published data.[37,38] When the NCP reaches zero, the cooler achieves its equilibrium temperature (ET). The selective emitter achieves a much lower ET than that of the broadband emitter, indicating a better sub-ambient cooling performance. For the two kinds of coolers exposed to an ambient temperature of 20°C, we present in **Fig. 1E** the NCP (when the cooler was at $T_{amb}$) and ET versus the PWV. When the PWV is smaller than 2 cm, the atmospheric transmissivity over the entire infrared spectrum is relatively high due to the little air moisture, resulting in higher NCP of the broadband cooler than that of the selective one. With increasing the PWV, the transmissivity of the spectrum out of the atmospheric window decreases significantly. As a result, the NCP of the two coolers decreases until they eventually become the same. For both coolers, the equilibrium temperature rises with increasing PWV, indicating a fading sub-ambient cooling performance. The selective cooler, owing to its highly selective spectrum, absorbs minimal thermal radiation from the atmosphere and consistently sustains a lower temperature compared to the broadband cooler. Results in **Fig. 1E** indicate the PWV significantly impedes the RC performance of a cooler, whether broadband or selective emitting. We also note that even at very high PWV, there is still a non-zero cooling power of 30 W·m$^{-2}$. Therefore, radiative cooling effects should be observable even in the scenarios with high humidity, in consistency with experimental observations[39].

## 3. Validation against experimental data

The reliability of the present results obtained via our proposed model (see Methods) was comprehensively verified. Firstly, the spectral irradiance from the hemispherical sky at PWV = 2.0 cm is obtained via the present model. It is plotted in **Fig. 2A** (see **Fig. S1** for comparison results at other PWV values) and compared against the benchmarks generated via MODTRAN. The average relative error of the modeled results is 2.4%, demonstrating that the proposed model in this work could accurately account the atmospheric transparency and its back radiation to the radiative cooler. Subsequently, the atmospheric back radiation was derived by integrating the spectral irradiance across all wavelengths. The atmospheric back radiation under different PWV conditions was calculated using the present model, as illustrated in **Fig. 2B**, and it is compared with the MOTRAN results. The average relative error is as small as 0.38%, indicating the accuracy of the proposed model for calculating the atmospheric back radiation to the radiative cooler under various PWV levels.

Furthermore, we compared the equilibrium temperature calculated by the present model with the experimental results. **Figure 2C** illustrates the solar irradiance and equilibrium temperature reduction in three locations: Maryland, USA (39 °N, 77 °W), Beijing, China (39 °N, 120 °E), and Suzhou, China (31 °N, 120 °E). Using experimental parameters in Maryland given by Zhao et al.[40], we calculated the equilibrium temperature distributions and compared them with published results. As sketched in **Fig. 2D**, the modeled results agreed well with the tested data. Based on the experi-

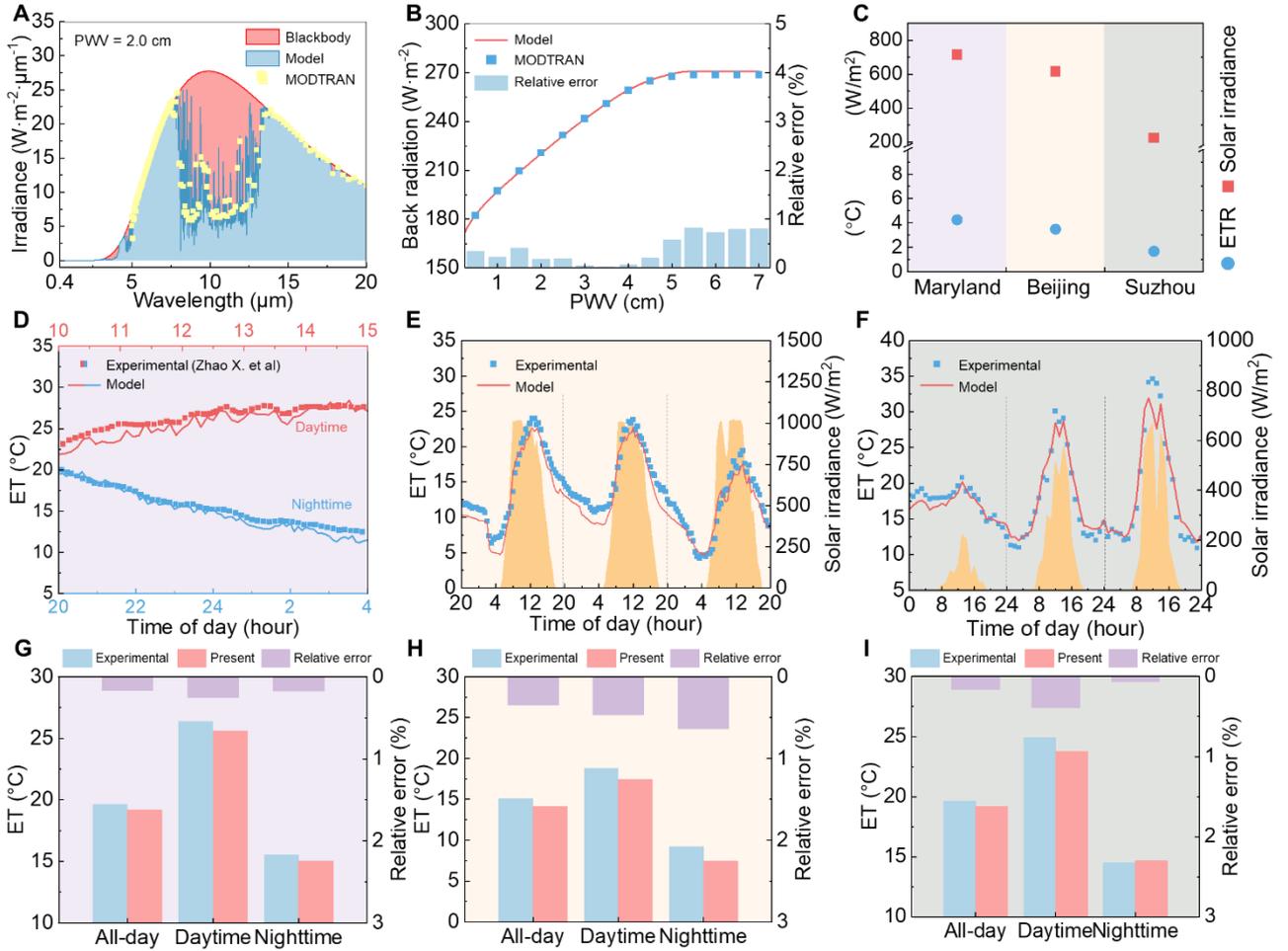

**Figure 2. Correctness validation of the present model for predicting radiative cooling capacity.**

(**A**) Spectral irradiance of the atmosphere at a precipitable water vapor (PWV) of 2.0 cm, obtained using our present model and compared with MODTRAN results.

(**B**) Atmospheric back radiation power at different PWVs, obtained using our present model and compared with MODTRAN results.

(**C**) Experimentally measured solar irradiance and equilibrium temperature reduction (ETR) from Maryland, USA by Zhao et al. [40], Beijing, China, and Suzhou, China.

(**D** to **F**) Comparison of equilibrium temperature (ET) obtained using our presented model and experimental data from (**D**) Maryland, USA by Zhao et al. [40], (**E**) Beijing, China, and (**F**) Suzhou, China.

(**G** to **I**) Comparison of average ET obtained using our presented model and experimental data from (**G**) Maryland, USA by Zhao et al. [40], (**H**) Beijing, China, and (**I**) Suzhou, China.

mental data, the average errors of the modeled results were 0.16%, 0.25%, and 0.17% (Kelvin temperatures were utilized for relative error assessment) for the all-day, daytime, and nighttime data, respectively, as plotted in **Fig. 2G**. Additional validations for other worldwide locations are presented in **Fig. S2**.

We also conducted field experimental tests of a radiative cooler (see **Fig. S3** for cooler's emissivity) located in the Chinese cities of Beijing (dry weather), Suzhou (moist weather), and Weihai (coastal weather), respectively (see **Figs. S4, S6, and S10** for experimental setups). We presented in **Fig. 2E** the uninterrupted three-day (see **Fig. S5** for seven-day results) temperature record of the cooler at Beijing. **Figure 2H** indicates the relative errors of 0.34%, 0.47%, and 0.64% of the modeled results from the experimental data, for the entire testing period, daytime, and nighttime, respectively. **Figure 2F** plots the temperature record of the cooler at Suzhou (see **Fig. S7** for 30-day data) and **Fig. 2I** shows the relative errors (0.16%, 0.39%, and 0.06% for the all-day, daytime, and nighttime data, respectively). In contrast, results obtained via MODTRAN show obvious larger deviations from the experimental data (see **Figs. S8** and **S9**). Temperature comparison for the coastal city of Weihai is presented in **Fig. S11**. Collectively, experimental results from our own and worldwide groups demonstrate the reliability of our assessment model for evaluating the radiative cooling capacity across different regions of the world.

## 4. Global radiative cooling capacity

Next, we investigate the NCP and equilibrium temperature reduction (ETR) achieved by the ideal broadband and selective radiative coolers to comprehensively evaluate the worldwide RC capability. It is worth mentioning that we first calculated hourly NCP and ETR from hourly meteorological and PWV data. Then, we obtained the monthly and annual average results by averaging the hourly NCP and ETR. When the cooler operates at the local ambient temperature, the average NCP during the northern hemispheric summer (June, July, and August), winter (November, December, and January), and the whole year are presented in **Figs. 3A**, **B**, and **C** for the broadband cooler and **Figs. 3D**, **E**, and **F** for the selective one (see **Figs. S16** and **S17** for monthly NCP distributions). The comparison between the broadband and selective coolers reveals that the former cooler exhibits superior performance. This is because when the coolers are fixed at the same ambient temperature, the broadband cooler can achieve additional cooling power through the secondary atmospheric window of $16 - 20$ μm. Overall, the NCP is relatively high near the latitude of 30° and low at the equatorial and polar areas. The explanations are as follows. The equatorial locations are usually at high humidity due to the intense evaporation of seawater, while locations at high latitudes are usually subjected to low temperatures and high cloud cover due to the polar climate. The area around the latitude of 30° is influenced by the subtropical high-pressure belt. The up-to-down airflow reduces water vapor condensation and the formation of a dry, cloudless sky environment. Consequently, the NCP of a radiative cooler is higher when it is located in areas with a latitude of around 30°, compared with the lower cooling power at the equatorial and polar regions.

During the northern hemispheric summer, it is seen from **Figs. 3A** and **D** that a NCP that surpasses $100$ W·m$^{-2}$ is achieved in the core area of the North American and Sahara deserts where the climate exhibits a scorching and arid environment. The regions with a radiative cooling power higher than $80$ W·m$^{-2}$ correspond to several desert regions, including the Arabian Desert (Arabia), the Sahara Desert (North Africa), the Kalahari Desert (South Africa), the Deserts of Australia (A to C) Maps of NCP achieved via the ideal broadband cooler during the Northern Hemisphere (A) summer, (B) winter, and (C) the whole year.

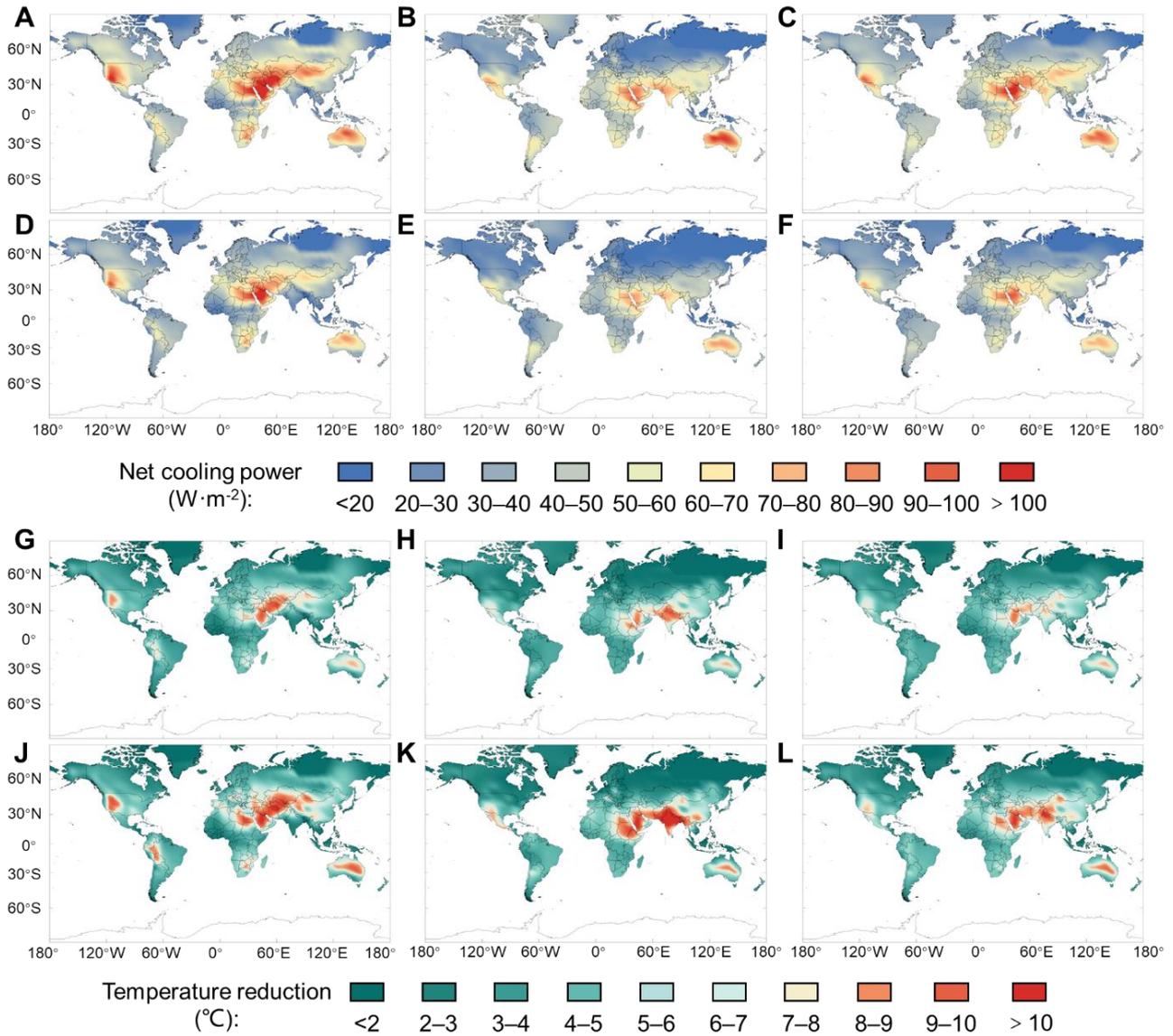

**Figure 3. Net cooling power (NCP) and equilibrium temperature reduction (ETR) via ideal broadband and selective coolers.**

(D to E) Maps of NCP achieved via the selective cooler during the Northern Hemisphere (D) summer, (E) winter, and (F) the whole year.

(G to I) Maps of ETR achieved via the broadband cooler during the Northern Hemisphere (G) summer, (H) winter, and (I) the whole year. (J to L) Maps of ETR achieved via the selective cooler during the Northern Hemisphere (J) summer, (K) winter, and (L) the whole year.

(Australia), and the Deserts of North America (North America). These desert regions are characterized by dry and cloudless climates, and the atmospheric transparency is usually at a high level. Consequently, a substantial net radiative cooling power can be achieved in these areas.

During the Northern Hemispheric winter, results in **Figs. 3B** and E reveal that the radiative cooler's performance is weakened by low ambient temperatures and cloudy days. Notably, the NCP in the east and southeast of Asia shows a slight improvement due to the weakening of the southeast monsoon in winter. Additionally, the Indian Peninsula is in the dry season, significantly enhancing net cooling power. In contrast, the Southern Hemisphere is experiencing a hot summer, and the larger ocean area brings more precipitation, leading to a decrease in NCP throughout the Southern Hemisphere, including two desert areas. **Figs. 3A**, **B**, **D**, and E indicate a consistent trend in the NCP across both hemispheres, with relatively high values in June, July, and August, and low values in November, December, and January, except for a small portion significantly influenced by monsoons. This trend is not influenced by the seasonal variations between the hemispheres.

As the extensive land area in the Northern Hemisphere leads to a limited influence of the ocean on the inland regions, the maximum of the NCP in the Northern Hemisphere is higher than that in the Southern Hemisphere, which can be seen from the year-averaged NCP presented in **Figs. 3C** and **F**. The NCP significantly decreases in both the northern and southern hemispheres as the distance to the poles shortens. Collectively, the radiative cooling capacity in the northern hemisphere is higher than that in the southern hemisphere.

The ETR achieved by a radiative cooler is presented in **Fig. 3**. The global maps of the temperature reduction achieved by a broadband cooler are presented in **Figs. 3G** and **H** for the northern hemispheric summer and winter, and **Fig. 3I** for the whole year. Those achieved by a selective cooler are presented in **Figs. 3J, K**, and **L** (see **Figs. S18** and **S19** for monthly ETR distributions). Typically, the cooler's equilibrium temperature is lower than the ambient temperature. As a cooler gains thermal intake from the atmosphere via radiation and surroundings via non-radiative parasitic heat loss [see Equation 21 in Methods for specific heat loss connefficients], the spectrally selective cooler absorbs less atmospheric radiation, and its temperature reduction surpasses, generally by 3 °C, that achieved via the broadband one. Specifically, in **Fig. 3J**, where the summer-averaged ETR achieved via a selective cooler is plotted, an ETR of more than 6 °C can be achieved within most areas between 20 °N and 40 °N, especially the areas of the North American Desert, Sahara Desert, Arabian Desert, and Karakum Desert. During the winter, the Indian Peninsula region experiences the highest season-averaged ETR, as shown in **Fig. 3K**, with over 10 °C observed across the vast majority of the area. This winter-averaged ETR surpasses even that observed during the summer. This phenomenon is attributed to that the region is experiencing its dry winter time, characterized by a dry atmosphere with high transparency and slower wind speeds compared to the summer months. These conditions contribute to a robust RC performance. For the year-averaged temperature reduction presented in **Figs. 3I** and **L**, an ETR over 4 °C and 5 °C is, respectively, achieved via the broadband and selective cooler operated at the area between 20 °N and 40 °N.

## 5. Quantification and classification

Results in **Fig. 3** indicate that the RC performance of a cooler significantly depends on the season and local climate features. In addition, the climate for different regions might exhibit different variations in the same year. Therefore, quantifying the year-round RC capacity at one location and classifying the capacity at different regions is important for implementing radiative cooling techniques. Here, we selected several representative cities, namely Beijing, China (39.8 °N, 116.5 °E), Mumbai, India (18.9 °N, 72.8 °E), Ankara, Turkey (40.1°N, 33.0°E), Asyut, Egypt (27.1°N, 31.0 °E), Kuala Lumpur, Malaysia (3.1 °N, 101.1 °E), Reconquista, Argentina (29.2 °S, 59.7 °W), Harare, Zimbabwe (17.9 °S, 31.3 °E) and Santiago, Chile (33.4 °S, 70.8 °W), to quantitatively analysis the monthly radiative cooling capacity and then categorize the global map into five zones with differentiable cooling performance.

The monthly NCP distributions achieved via the ideal broadband cooler placed in the cities mentioned above are depicted in **Fig. 4A** [see **Tables S1** to **S4** for detailed data, **Fig. S20** for results obtained using the conventional model that dismisses the varying atmospheric transparency (our model can capture the monthly changing trend while the conventional model failed), and **Fig. S21** for monthly energy saving capacities]. It is seen that in the city of Ankara, the NCP during the local summer months, June, July, and August, is higher than that during its winter months of November, December, and January. Similarly, in Santiago, the NCP during the local summer months, November, December, and January, is larger than that during the local winter months of June, July, and August. These two cities are representatives of regions with high NCP during the local summer and low NCP during the local winter in the northern and southern hemispheres, respectively. Such regions typically experience the so-called Mediterranean climate, characterized by high temperatures and dry air in summer, while low temperatures and moist air in winter. These climate features allow strong RC performance in the summer when cooling power is in high demand and low NCP in winter when RC is supposed to be avoided. The NCP of a radiative cooler in such regions can well match the seasonal cooling demand. Therefore, radiative coolers are recommended in such regions.

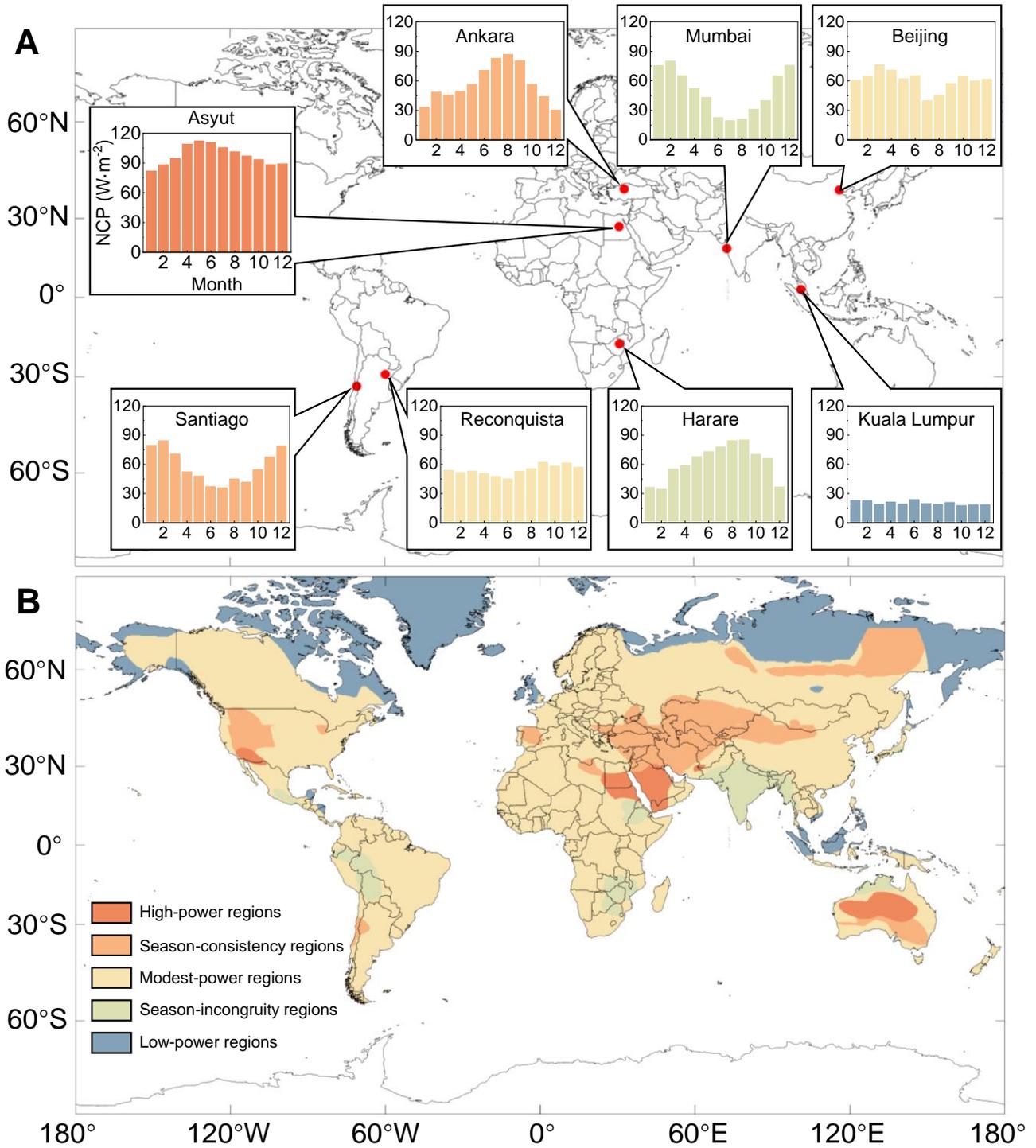

**Figure 4. Quantification and classification of the global radiative cooling capacity.**

(A) Monthly NCP distributions for typical Northern and Southern Hemisphere cities.

(B) Classified radiative cooling regions on Earth.

Contrary to the situation in Ankara, the mean NCP during Mumbai's summer months, June, July, and August, is apparently lower than that during its winter months of November, December, and January. In Harare, the NCP during the local summer months (November, December, and January) is also lower than that during the winter months (June, July, and August). Such abnormal phenomena are typically induced by the monsoon. In summer, the monsoon rainfall is heavy, and the strong PWV suppresses the NCP of a radiative cooler. When it comes to winter, the monsoon effect becomes weak, and the dry atmospheric environment facilitates substantial NCP. These two cities represent regions in the northern and southern hemispheres, respectively, where the NCP trends to be higher NCP during the local winter and lower in summer.

In Beijing and Reconquista, a cooler's monthly NCP is substantial, around 60 $W \cdot m^{-2}$, and exhibits relatively tiny fluctuations. This stability is because such cities generally experience minimal changes in monthly precipitation and maintain a stable environment over a year. These two cities represent the regions with a stable year-round NCP in the northern and Southern Hemispheres, respectively. In Asyut, a radiative cooler exhibits a relatively high monthly NCP, above 86 $W \cdot m^{-2}$ with a maximum of 118 $W \cdot m^{-2}$, representing regions with annual stable and high cooling capacity. Such regions typically have a desert climate, with high temperatures, dry air, and less cloud cover, which boosts the radiative cooling performance. On the other hand, Kuala Lumpur exhibits a low monthly NCP with a maximum of 25 $W \cdot m^{-2}$ throughout the year, representing a moronic RC region with high precipitation and cloud cover, which impede the dissipation of exhaust thermal radiation.

Based on the qualitative analysis in **Fig. 3** and the quantitative analysis in **Fig. 4A**, we classified the global map of RC power into five distinct regions, namely, high-power, season-consistency, modest-power, season-incongruity, and low-power regions, as illustrated in **Fig 4B** (see **Fig. S22** annual energy saving capacity). The high-power regions typically have a desert climate and are mainly located in North Africa and Australia. These regions are well-suited for large-scale implementation of RC technologies and facilities due to the all-year high temperatures, low PWV-induced transparent atmosphere, and low population density. Besides, promoting RC implementations in these areas is expected to mitigate global warming in limited regions. The season-consistency region mainly includes the littoral areas of the Mediterranean and the west coast of South and North America. They share similar atmospheric moisture characteristics that match the seasonal cooling demand, resulting in high summery and low hiemal cooling power. Therefore, a radiative cooler in such regions can exhibit high performance for thermoregulation and maximize the annual energy saving. The modest-power regions fill the majority of the landscape around the world, as seen in **Fig. 4B**. These regions lack the seasonal shift between rainy and dry climates, maintaining a relatively stable atmospheric characteristic throughout the year. Consequently, the monthly RC power in these areas is moderate with minimal fluctuations. In regions experiencing seasonal incongruity, the NCP is low during the local summer and high in the winter, primarily influenced by the monsoon climate. This climate pattern results in a moist atmosphere during summer and a dry one during winter. These regions are primarily distributed in the Indian Peninsula, the east coast of Africa, the west coast of South America, and northern Australia. Large-scale RC implementation in the season-incongruity regions may result in relatively low benefits, so it is necessary to carefully consider the investment cost and the resulting returns. The low-power regions are found to be located in two areas. The first one is the equatorial tropic area where the PWV is too strong that a radiative cooler therein is basically failing to gain useable NCP. The other low-powered area fills the Earth's poles where the low temperature and substantial cloud cover impede the NCP of a cooler.

## 6. Conclusions

In summary, the RC capacity on Earth, at the season- and year-averaged level, is quantitatively assessed by using recorded meteorological data. The cooling performance map indicates that compared with the Southern Hemisphere area, the Northern Hemisphere exhibits better RC capacity because of its year-round match between the climate characteristics and radiative cooling demands. Further, the Earth's surface is classified into five representative regions based on the monthly RC capacity. Among the five regions, the season-consistency region performs well for annual energy saving and is recommended for large-scale RC applications. Inversely, the season-incongruity region shows a mismatched RC power with cooling demand. The high-power region mainly contains desert areas where large-scale RC applications pave a promising way for alleviating global warming. The low-power region is mainly located at the Earth's poles and a limited Equatorial region. The major part of the Earth's face belongs to the modest-power region where the NCP is appreciable and remains stable throughout the year. Therefore, research on dynamic RC via self-regulated emissivity or reflectivity of a cooler shows enormous potentiality for passive energy harvesting from space in the power-modest region. This work paves a reliable tool for predicting global RC capacity and guides worldwide applications of RC-based technologies.

**Methods**

In this section, variables that do not specify units use Standard International Units.

**The precipitable water value model**

The meteorological data used in this work was obtained from Energyplus. First, ground meteorological data are used to calculate PWV by using the basic theory of GPS-based PWV estimation.

The delay in GPS signals induced by the troposphere is often assessed using the zenith total delay (ZTD), representing an equivalent excess path length. The Saastamoinen model[41] is commonly employed to model this delay, suggesting a division of the troposphere into two distinct layers. The first layer extends from the ground to a height of 10 km and exhibits a temperature decrease of 6.5 °C·km$^{-1}$. The second layer spans from 10 km to the top of the troposphere at 70 km, and the temperature remains constant within this layer. Therefore, when integrating the refractive index of the atmosphere, the refractive index function can be expanded according to the trigonometric function of zenith distance and the termwise integration. The ZTD (in cm) calculated by this method is[41]

$$ZTD = 0.2277 \times \frac{\left[ P_s + \left( 0.05 + \frac{1255}{T_s} \right) e_0 \right]}{f(\varphi, h)},$$ (Equation 1)

Here, $P_s$ is the Earth's surface pressure (in hPa). $T_s$ is the Earth's surface temperature. $e_0$ is the Earth's surface water vapor pressure and can be calculated by[42]

$$e_0 = rh \times 6.11 \times 10^{\frac{7.5 \times (T_s - 273.15)}{T_s}},$$ (Equation 2)

with $rh$ denoting relative humidity. $f(\varphi, h)$ is the correction of gravitational acceleration caused by the Earth's rotation and can be written as[42]

$$f(\varphi, h) = 1 - 0.00266\cos 2\varphi - 0.00028h, \qquad \text{(Equation 3)}$$

where $\varphi$ is latitude, and $h$ is altitude (in km).

The ZTD can be decomposed into two distinct components: the zenith hydrostatic delay (ZHD) and zenith wet delay (ZWD)[41]. The ZHD primarily arises due to dry gases in the atmosphere and typically constitutes around 90-100% of the total delay, with the remaining portion attributed to the wet delay. The model proposed by Elgered et al.[43] provides an accurate method to estimate ZHD (in cm). It relies on utilizing surface pressure data and assumes that the atmosphere is in a hydrostatic equilibrium state, which enables a reliable representation of the ZHD component[44]

$$\text{ZHD} = \left(0.22779 \pm 0.00024\right) \frac{P_s}{f(\varphi, h)}, \qquad \text{(Equation 4)}$$

The ZWD is calculated from ZTD and ZHD[44]

$$\text{ZWD} = \text{ZTD} - \text{ZHD}, \qquad \text{(Equation 5)}$$

Then, the PWV can be obtained via ZWD by using[44]

$$\text{PWV} = \Pi \times \text{ZWD}, \qquad \text{(Equation 6)}$$

where $\Pi$ is the conversion factor and can be calculated by[44]

$$\Pi = \frac{10^6}{\rho R_v \left[ k_2 + \dfrac{k_3}{T_m} \right]}, \qquad \text{(Equation 7)}$$

where $\rho = 999.97$ is the density of liquid water, $R_v = 461.51$ is the specific gas constant of water vapor, $k_2 = 22.1 \pm 2.2$ (K·hPa$^{-1}$), and $k_3 = 373900 \pm 1200$ (K$^2$·hPa$^{-1}$) are atmospheric refraction constants[45], $T_m$ is the weighted average temperature of the atmosphere, conventionally obtained via[35]

$$T_m = 70.2 + 0.72 T_{amb}, \qquad \text{(Equation 8)}$$

Then, the PWV calculated from Eq (6) is used as a variable to generate the spectral atmospheric emissivity by using MODTRAN[36]. Finally, we utilized hourly weather data from Energyplus to calculate the hourly NCP and ETR (Number of data: $24 \times 365 = 8760$), and then obtained their monthly and annual averages. The map is generated using triangulation-based cubic interpolation. The modeled NCP is compared against experimental results in **Fig. 2** and **Fig. S2**, which confirms the correctness of the results presented in this article.

**Radiative cooling energy balance model**

The NCP of a cooler, $P_{net}(T_e, T_{amb})$, is written as[2]

$$P_{net}(T_e, T_{amb}) = P_{rad}(T_e) - P_{atm}(T_{amb}) - P_{sun} - P_{loss},$$ (Equation 9)

where $T_e$ is the temperature of the cooler, and $T_{amb}$ is the ambient temperature.

In Eq (9), $P_{rad}(T_e)$ is the power radiated by the cooler and is written as[2]

$$P_{rad}(T_e) = A \int d\Omega \cos\theta \int_0^\infty d\lambda I_{BB}(T_e, \lambda)\varepsilon(\lambda, \theta),$$ (Equation 10)

The atmospheric back radiation power absorbed by the cooler, $P_{atm}(T_{amb})$, can be calculated via[2]

$$P_{atm}(T_{amb}) = A \int d\Omega \cos\theta \int_0^\infty d\lambda \varepsilon(\lambda, \theta) I_{BB,atm}(T_{amb}, \lambda)\varepsilon_{atm}(\lambda, \theta),$$ (Equation 11)

Here, $A$ and $\varepsilon(\lambda, \theta)$ are the sky-facing area and emissivity of the cooler, respectively. The atmospheric emissivity $\varepsilon_{atm}(\lambda, \theta)$ in the polar angle $\theta$ can be calculated as[2]

$$\varepsilon_{atm}(\lambda, \theta) = 1 - t_{atm}(\lambda)^{1/\cos\theta},$$ (Equation 12)

where $t_{atm}(\lambda)$ denotes the atmospheric transmittance in the vertical direction, and $I_{BB}(T, \lambda)$ is the radiation intensity of a blackbody at wavelength $\lambda$ and temperature $T$, written as

$$I_{BB}(T, \lambda) = \frac{2h_{plk}c^2 / \lambda^5}{e^{h_{plk}c/(\lambda k_B T)} - 1},$$ (Equation 13)

with $h_{plk}$ denoting the Plank constant, $k_B$ the Boltzmann constant, and $c$ the light speed.

Among the primary absorbing media in the atmosphere, ozone is mainly distributed in the stratosphere. The temperature of the stratosphere is ~220K, significantly lower than the ambient temperature[46]. If the ambient temperature is used to calculate atmospheric irradiance, it would lead to an overestimation of the results. Therefore, the estimation formula for atmospheric irradiance $I_{BB,atm}(T,\lambda)\varepsilon_{atm}(\lambda, \theta)$ is modified as[47]

$$I_{BB,atm}(T,\lambda)\varepsilon_{atm}(\lambda,\theta) = I_{BB,rest}(T_{amb},\lambda)\varepsilon_{rest}(\lambda,\theta) + I_{BB,ozone}(T_{ozone},\lambda)\varepsilon_{ozone}(\lambda,\theta) \,, \quad \text{(Equation 14)}$$

Here, each component's emissivity $\varepsilon_i(\lambda, \theta)$ is calculated by using **Eq. (12)**. The transmittance of ozone is calculated via[47]

$$t_{ozone}(\lambda) = \frac{t_{atm}(\lambda)}{t_{rest}(\lambda)}. \quad \text{(Equation 15)}$$

The atmospheric irradiance obtained using the present model compared against the results using the conventional model are shown in **Figs. S12** and **S13**. The atmospheric back radiation, $P_{atm}(T_{amb})$, calculated via **Eq. (11)**, is only suitable for a clear sky without any clouds. In actual weather conditions, the influence of cloud cover is non-negligible and must be considered. In meteorological observations, the cloud cover, denoted by $C_c$, which is the area ratio of clouds that occupy the sky and ranges from 0 to 1[48], is used to quantify the cloud condition. Therefore, the realistic back atmospheric radiation that applies to all sky cases reads as[30]

$$P_{atm,C_c}(T_{amb}) = A\int d\Omega\cos\theta \int \left(I_{BB}(\lambda,T_{amb})\varepsilon_{atm}(\theta,\lambda)(1-C_c) \right. \\ \left. + I_{BB}(\lambda,T_{cloud})C_c\right)\varepsilon(\theta,\lambda)\,d\lambda \,, \quad \text{(Equation 16)}$$

Here, $T_{cloud}$ is the cloud temperature (see **Figs. S14** and **S15** for cloud temperature influences on atmospheric back radiation). For the cloud with a base height of $H_{cb}$, its temperature can be calculated via[41]

$$T_{cloud} = T_{amb} - \gamma H_{cb},$$ (Equation 17)

where $\gamma$ is the atmospheric lapse rate of 6.5 °C·km$^{-1}$. The cloud base height can be estimated through the lifting condensation level and can be expressed as[49,50]

$$H_{cb} = 0.124 \times \left( T_{amb} - T_{dew} \right),$$ (Equation 18)

where $T_{dew}$ is the local dew point temperature.

The term of $P_{sun}$ in **Eq. (9)** is the incoming solar radiation absorbed by the cooler and can be written as[2]

$$P_{sun} = A \int_0^\infty d\lambda \varepsilon(\lambda, \theta) I_{sun}(\lambda),$$ (Equation 19)

The intensity $I_{sun}(\lambda)$ is the solar radiation that reaches the cooler. The non-radiative parasitic heat loss induced by the conduction and convection between the cooler and the ambient air is expressed as[29]

$$P_{loss} = hA(T_{amb} - T_e),$$ (Equation 20)

where $h$ is the non-radiative heat transfer coefficient and can be calculated by[29]

$$h = 2.5 + 2v,$$ (Equation 21)

Here, $v$ is the wind speed.

**Declaration of Competing Interest**

The authors declared that they have no conflicts of interest with this work.

**Data availability**

The data presented in this study are available from the lead contact (C. Wang) upon reasonable request.

**Acknowledgments**

This work was supported by the National Natural Science Foundation of China (51906014).